**Aldo Fabricio Martínez Fiorenzano**

**Università degli Studi di Padova**
**INAF - Osservatorio Astronomico di Padova**

**PhD Thesis in Astronomy**

**The search for extrasolar planets:**
**Study of line bisectors and its relation**
**with precise radial velocity measurements**

March 27th, 2006.
Padova, Italy

## Abstract

In recent years, the study of the mechanisms of formation and evolution of planetary systems has received a considerable boost from the discovery of more than a hundred extra-solar planets, mainly thanks to the analysis of the variations of radial velocities of the stars. While several general features of planetary systems are beginning to emerge, still little is known of several aspects, concerning e.g. the possible mechanisms that lead to the observed planet configurations (semimajor axis, orbital eccentricity, planetary masses, etc.). In particular, the impact of dynamical interactions in wide binary systems (a very common case among stars in the solar neighborhood) is still unknown.

This has significant impact on e.g., the determination of the frequency of planets in general, and of those able to host life in particular.

With the aim to contribute to this field, a long term program has begun at INAF using the "Telescopio Nazionale Galileo" (TNG) on a sample of about 50 wide binary systems. The program searches for Jupiter-sized planets in these systems using variations of the radial velocities. A few detections would be expected, based on statistics for single stars. However, radial velocity variations of stars due to planets are small, typically of the order of a few tens of m/s, or even less.

Apparent variations of similar size can be caused by effects other than Keplerian motion of the stellar barycentre. The purpose of this study is to develop a technique able to distinguish between radial velocity variations due to planets from the spurious variations due to stellar activity or spectral contamination, with the aim to search for planets around young/active stars, and to clean our sample from possible erroneous measures of radial velocities.

To this purpose, in the course of the thesis work we prepared a suitable software in order to use the same spectra acquired for radial velocity determinations (i.e., with the spectrum of the Iodine cell imprinted on) to measure variations of the stellar line profiles. This is a novel approach, that can be of general utility in all high precision radial velocity surveys based on iodine cell data. This software has then been extensively used on data acquired within our survey, allowing a proper insight into a number of interesting cases, where spurious estimates of the radial velocities due to activity or contamination by light from the companions were revealed. The same technique can also be considered to correct the measured radial velocities, in order to search for planets around active stars.

The structure of the thesis is as follows. In Chapter 2 some general aspects about ongoing situation in the research field of extrasolar planets are exposed. Current theories about planet formation, like the core accretion and the disk instability, as well as proposed mechanisms of planet migration to explain the presence of massive planets in very close orbits around their host stars, are briefly presented and commented.

In Chapter 3, various detection techniques are described, with special emphasis on the Doppler or radial velocity technique, and the two methods employed for high precision measurements, through the Iodine cell and the simultaneous wavelength calibration with optical fiber fed spectrographs, are discussed.

Relevant aspects of stellar atmospheres are presented in Chapter 4, with a brief description of stellar activity and of the usefulness of line bisectors in the interpretation of physical processes through the study of spectral line asymmetries.

In Chapter 5, we present the current status of the Italian planet search program around wide binaries, ongoing at the TNG, with the radial velocity technique employing the Iodine cell with the high resolution spectrograph SARG. Some characteristics of the stellar sample, results, and future perspectives are given.

We developed a software able to read and analyze the stellar spectra with the Iodine lines. A description of the technique employed to remove the Iodine features from the stellar spectrum is given in

Chapter 6: it exploits the spectrum of a rapidly rotating B-star spectrum, acquired within the same procedure adopted to measure precise radial velocities. This allows to deal with spectra free of Iodine lines to perform a detailed analysis of spectral line asymmetries. A solar catalog was employed to construct a mask, which is cross-correlated with the stellar spectra to obtain high S/N average absorption profiles; these were used to measure line bisectors (i.e. the middle point at constant flux between the blue and red sides of the profile). The constancy in time of the shape and orientation of line bisectors would ensure that radial velocity variations measured for a star are due to the barycentre motion, caused by a substellar companion orbiting the observed star. The difference of velocities given by an upper and lower zone of the line bisectors, known as bisector velocity span, is employed in a plot against radial velocities to search for possible trends and thus correlations. Outliers (mainly due to contamination by light from companions) can also be identified on these plots.

In Chapter 7, the analysis of a subsample of the program stars is presented. Details about the chosen subsample and the motivations for the choice of upper and lower zones to determine the bisector velocity span in the search for possible correlations, are given. The instrumental profile characteristics are described, and its (negligible) influence on the asymmetries observed in the stellar spectra is discussed. Some statistical results are also presented.

In Chapter 8, the meaning of the correlations are discussed and explained for the specific cases of active stars and for the cases of the stellar spectra contaminated by light from a nearby object. A linear correlation with negative slope is found in the case of active stars, while for stars with their spectra contaminated by light from their companions the correlation is positive. For stars known to host planets, no correlation is found and line bisectors appear constant.

Finally in Chapter 9 we explored the possibility to apply corrections to the observed radial velocities in the case of stellar activity. These corrected radial velocities may be used to search for orbital motion, hidden by the activity variations, and/or to derive more stringent upper limits to possible substellar companions by Monte Carlo simulations. The success of such correction technique is discussed, as well as its usefulness in surveys looking for planets around young and active stars. Due to the intrinsic brightness of young planets, these represent important targets for direct imaging instruments.